# FOG AND DEW COLLECTION PROJECTS IN CROATIA

M. Mileta*
Metorological and Hydrological Institiute of Croatia
Zagreb, Croatia

D. Beysens*, V. Nikolayev*
CEA-Grenoble, and ESPCI, Paris, France)

I. Milimouk*
ESPCI
Paris, France

O. Clus*, M. Muselli*
Université de Corse and CNRS
Ajaccio, France

*and International Organization for Dew Utilization OPUR
Paris, France

## Abstract

*The present paper discusses the fog and dew water collection in Croatia. Zavižan, the highest meteorological station in Croatia( 1594m) is chosen for collecting of fog water with a standard fog collector (SFC). The highest daily collection rate was 27.8 L / m². The highest daily collection rate in days without rain was 19.1 l/m².*

*Dew is also a noticeable source of water, especially during the drier summer season. Dew condensers in Croatia have been installed on the Adriatic coast (Zadar) and islands Vis and Biševo. We report and discuss the data collected since 2003. In the small Biševo island, a special roof has been designed to improve the formation and collection of dew on a house. Data from April 2005 will be presented and discussed.*

*Keywords: dew collection, fog collection, water production, radiative condenser, Adriatic sea, Velebit*

## Introduction

This work concerns the dew and fog collection in Croatia. The objective was to estimate the amount of dew and fog water that can be collected, with emphasis on the dry summer season, where such alternate source of water can be the most helpful. Fog was studied on Velebit mountain at elevation 1594 m, and dew was studied at sea level on the Adriatic coast (Zadar) and islands Vis and Biševo.

## Fog water collection

Fog, from meteorological point of view, occurs when horizontal visibility is less than one kilometer, due to the presence of drops of water in the air. Use of fog as water resource is a known practice in the world. This is done by artificially collecting its water. Fog water collection in Croatia started with the Grunow type collector in 1954. In the summer of 2000 a new fog collector (SFC) was installed in Zavizan. The first results of fog water collected by SFC and the measurement by Grunow type collector were presented by Mileta (1998, 2003).

This paper discusses the daily fog water amounts collected with this new collector in 2000-2004 during the warm seasons.

## Site and methodology

Zavizan, the highest meteorological station in Croatia is located on Northern Velebit, 1594 meters above sea level (44°49'N, 14° 59'E). It is equipped with standard fog collector (SFC) for collecting fog water. The methodology used has been described in Schemenauer and Cereceda (1994) and is based on the used of a standard fog collector (SFC) of 1 m² of polypropylene mesh.

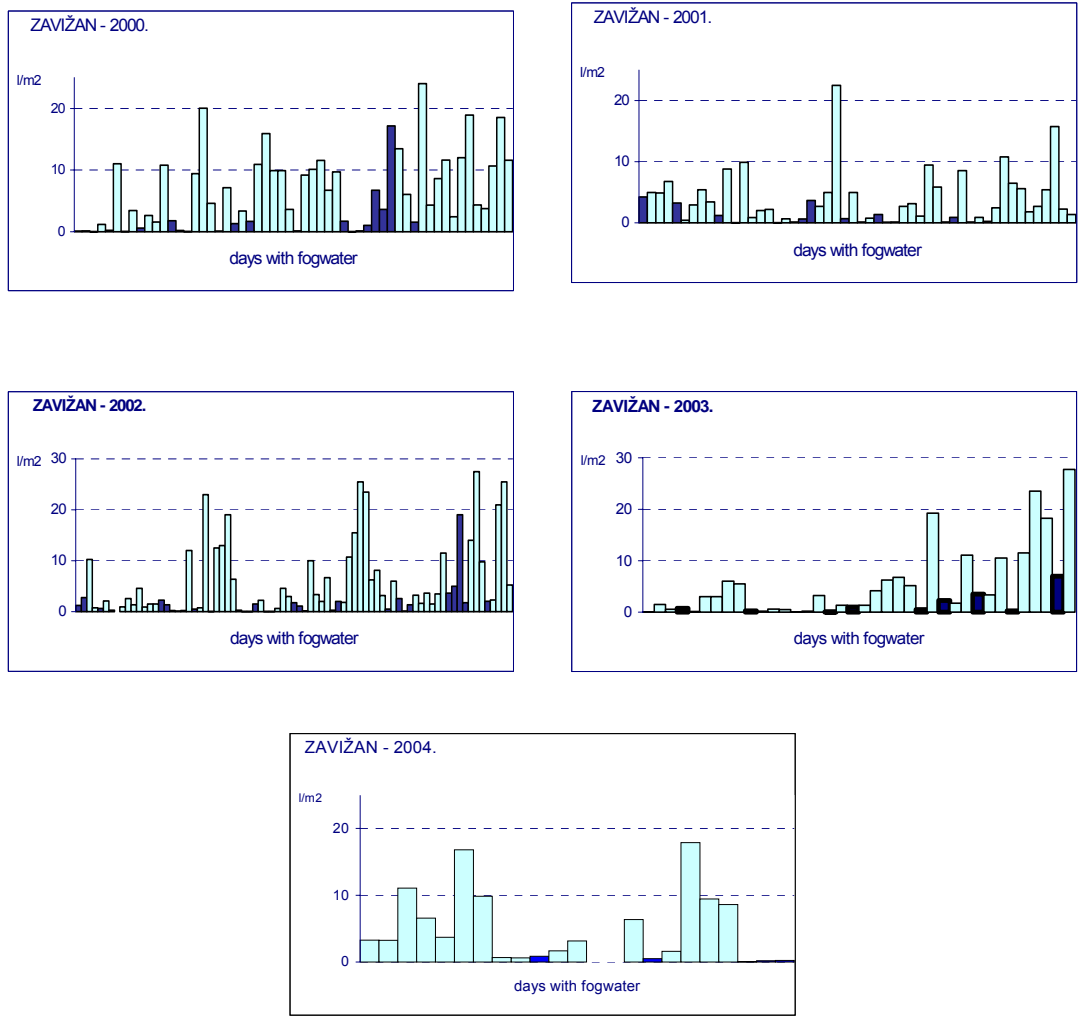

**Figure 1**. The daily amounts of collected fog water (2000-2004).

## Fog water results

The results presented in Figure 1 are the daily fog water amounts collected during warm seasons in 2000-2004. The data were obtained between July 27 and November 10 in 2000, May 16 and September 27 in 2001, June 26 and October 24 in 2002, July 3 and October 10 in 2003 and during June and September in 2004.
Maximum daily collected fog water was in the end of October and November, months with maximum days with fog (data of 2001 and 2003). The maximum one-day value was 27.8 l/m² on October 8,

2003. The highest daily collection rate in days without rain was 19.0 l/m² on October 16, 2002 (dark columns).

## Dew collection in Adriatic coast and islands

Dew is atmospheric vapour that condenses on a surface thanks to natural radiative cooling (Monteith, 1957; Beysens, 1995; Nikolayev et al., 1996]. Dew can bring substantial amount of water when the other resources (groundwater, rain, fog) are lacking. This is especially the case in the hot, dry season in the Mediterranean basin. In this paper, we investigate the dew water volume that can be harvested on materials that enhance natural (i.e. radiation driven) dew water condensation. We compare the dew yields obtained for different sites (coastal: Zadar, islands: Komiža and Biševo) by the Adriatic sea, in view of water recovery. The period of analysis was summer 2003 – summer 2004 for Zadar and Komiža, and summer 2005 for Biševo. Note that summer 2003 was exceptionally dry and hot in Western Europe.

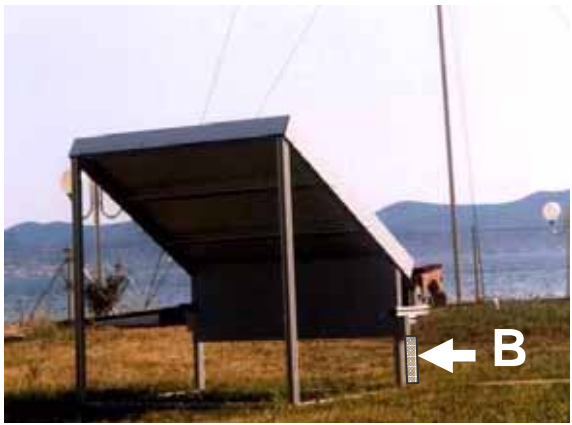

**Figure 2**. The dew condenser at the Zadar station (Croatia). B is the bottle for dew volume measurement.

### Sites

*Zadar* (Croatia). This station is located at latitude 44°08' N and longitude 15°13' E, 5 m a.s.l., on the Adriatic coast, north of Zadar, on a peninsula. The direction of the dominant winds during the night is NE. Windspeed is measured at 10m elevation. Data were collected between June 21, 2003 and May 31, 2004 (344 days).
*Komiža* (Vis island, Croatia). The island of Vis is in south-eastern Croatia, in the Adriatic Sea, within 45 km from the city of Split. It is the furthest island from the group of Central Dalmatian islands. Its surface area is 90.3 km². The station is situated at latitude 43°03' N, longitude 16°06' E, 20 m a.s.l. and dominant wind is SW. The site is situated in the middle of a mountainous cirque area open to the sea, with mountains of about 400 m. Wind speed is determined from the observation of the sea and is reported in Beaufort. Data were collected between June 24, 2003 and July 26, 2004 (399 days).
*Biševo*. This small island (5.84 km², maximum elevation 65 m asl) is located 20 km SW of Vis Island. The permanent population of this island is about 14 people in winter and 40 in summer. Biševo is deprived of natural water sources. The measurement site is located on the north side of the Salbunara bay, NW of Biševo island at latitude 42°56' N , longitude 16°47' E. Wind speed is measured at 1 m above the top of the roof and 3 m above ground. Data presented here were collected between April 22, 2005 and July 25, 2005.

### Experimental setup

In Zadar and Komiža, dew yield was measured on the same condensers of 1 m x 1 m inclined at a 30° angle from horizontal (Fig. 2). The angle of 30° is the "best" inclination angle for dew harvesting (Beysens et al. 2003). Both condensers are coated with a condensing foil made of $TiO_2$ and $BaSO_4$ micro spheres embedded in low density polyethylene (similar to Nilsson, 1996; made by OPUR, France, www.opur.u-bordeaux.fr) with hydrophilic surface. The foil is thermally insulated from the condenser frame by a 30 mm thick Styrofoam plate. Dew quantities were measured daily in the morning, corresponding to water collected by gravity flow in a bottle and scraped from the surface.

In Biševo, a 15.1 m$^2$ roof of a small house was covered with thermally isolated polycarbonate plates (Fig. 3) following the technique described by Muselli et al., 2002. The materials surface is treated hydrophilic on one side in order to condense a liquid film instead of droplets (treatment performed initially on the interior side, the material was turned upside down to expose the hydrophilic surface towards the sky). Dew water was *not* scraped.

In all sites, the following parameters were recorded on a regular basis: air temperature $T_a$ (°C), relative humidity $H$ (%), wind speed $V$ (Zadar: at 10 m elevation, in m/s; Biševo: in Beaufort; Komiža : at 1m above the roof, 3 m above the ground, in m/s), wind direction (degree), cloud cover data $N$ (Zadar and Komiža : octas). These parameters are recorded at the following times: Zadar (UTC +1): 08:00, 19:00, 21:00; Komiža (UTC +1): 07:00, 21:00; Biševo (UTC +2): every 15 min. Rain and frost were excluded from the dataset. In order to correlate with standard measurements at 10 m above the ground, the Biševo data are corrected by using the classical logarithmic variation (see e.g. Monteith and Unsworth, 1990):

$$V(z) = V_{10} \ln(z/z_c)/\ln(10/z_c), \quad (1)$$

where $z_c$ is the roughness length (measured in m) and taken equal to 0.1 m. This value corresponds to an open landscape obstacle of height, $H$, separated by at least 15 $H$. Eq. 2 gives $V(z = 3m) = 0.74\ V(z = 10m)$. The data are locally stored on a computer. The computer is connected to a GSM modem that permits a real time access to the data.

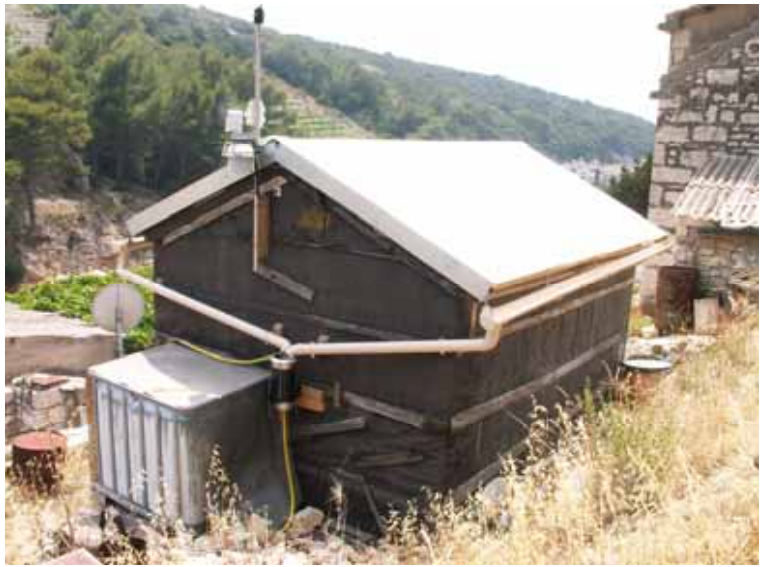

**Figure 3.** The house in Salbunara bay, NW of Biševo Island. The 15.1 m$^2$ roof has its symmetry axis oriented east (left)-west (right). Sea is on the west side. (Photo D. Beysens).

## Dew yields

Fig. 4 shows the daily pattern of the dew yield and Table 1 the average values. Seasonal variations can be observed. In Zadar, the winter dew yield is lower than in the other seasons, although it is in this season that the largest yields are observed in Komiža. Humidity and night duration increase in winter, thus the Zadar result is paradoxical. The reasons are due in part to (i) cloud coverage, greatest in winter, thus reducing the cooling power, (ii) strong wind speed in winter, preventing dew formation, (iii) data do not account for frost, (iv) local characteristics: Komiža, situated in the centre of a mountainous cirque open to the sea on the SW, with strong infrared scattering from the mountains heated by the summer sun, hindering dew formation during the hot season. The mean annual dew yields range is 0.08 mm for Komiža ) and 0.15 mm for Zadar. The percentage of dew days ranges from nearly 56 % (Biševo) to 19 % (Komiža). The accumulated dew yield during about 1-year period is 6.1 mm for Komiža and 13.3 mm for Zadar. Biševo should give a much larger yield as for about ¼ year, the cumulated yield is already 11.8 mm.

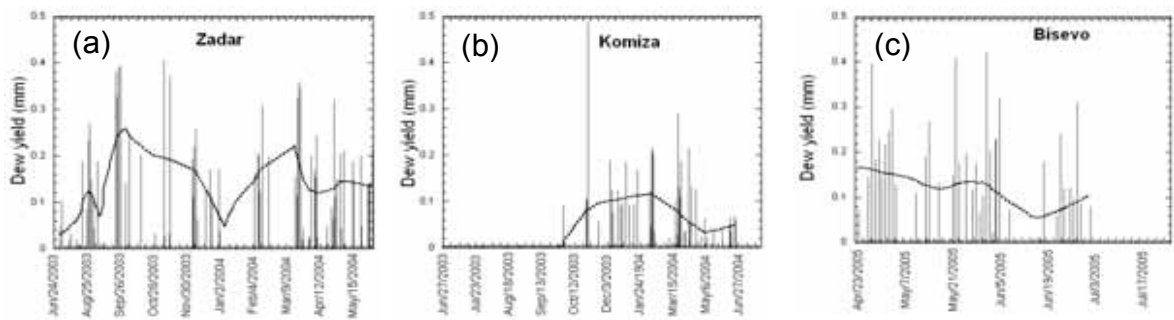

**Figure 4**. Daily dew collection: (a) Zadar, (b) Komiža : June 2003 – June 2004 (c) Biševo April-July 2005 (mm). The broken line is a smoothen fit of the data.

| Site | Zadar | Komiža | Biševo |
|---|---|---|---|
| dates (# of days) | June 21 2003 May 31 2004 (344) | June 24 2003 July 26 2004 (399) | April 22 2005 July 25 2005 (95) |
| dew days | 87 | 76 | 56 |
| dew days (%) | 25 | 19 | 59 |
| mean dew yield (mm) | 0.15 | 0.08 | 0.125 |
| Cumulated dew (mm) | 13.3 (about 1 year) | 6.1 (about 1 year) | 11.8 (about 1/4 year) |
| mean dew cloud cover (octas) | 3.1 | 2.9 | No data |
| mean dew wind-speed (m/s) | 1.7 | 1.7 – 3.1 (2 Beaufort) | 0.85 |
| mean dew RH (%) | 77.8 | 79.0 | 73.75 |

**Table 1** Mean values from the dew sites.

## Wind influence

Figure 5 compares the wind measured in the morning for 3 sites during dew events (see the experimental setup section for measurement time; data taken at 05:00 (UT+2) have been considered as representative for Biševo). The Beaufort scale has been translated in the upper scale of Fig.5b in wind speed classes (in m/s). There are no striking differences between the sites, which all show larger dew yields for lower wind speeds, with an upper limit of order 4 m/s at 10 m. The average wind speeds where dew forms correspond to rather large speed (1-2 m/s), an order of magnitude higher than found in continental locations, of order 0.1 - 0.2 m/s (Beysens et al. 2005).

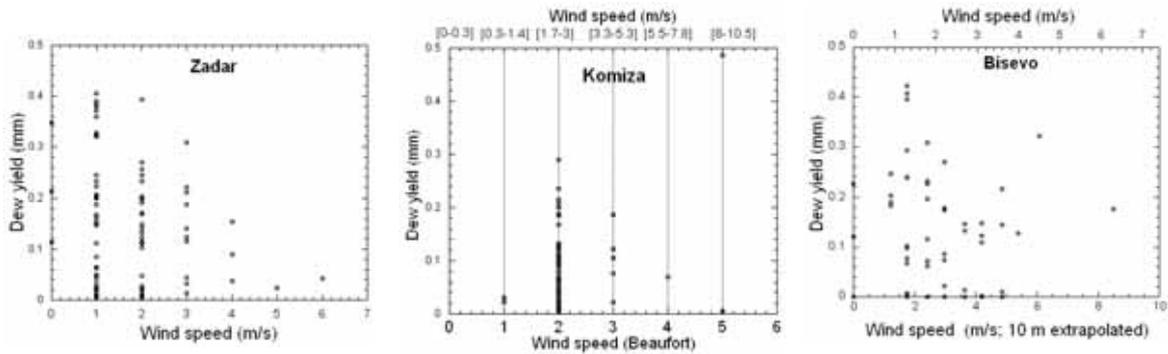

Figure 5. Dew yield with respect to wind speed at (a) Zadar; (b) Komiža . Here the Beaufort scale has been translated in wind speed classes in the upper scale; (c) Biševo, with the lower scale extrapolated at 10 m elevation according to Eq.1.

## Cloud cover

Figure 6 contains cloud cover data (in octas) at the time of dew collection. Oscillations in phase with the seasons were found (not shown), with, as expected, the highest cloud cover in winter. Dew forms with a mean cloud cover of 3 in both Zadar and Komiža sites. Dew yields are seen to rise when cloud cover diminishes because the radiative cooling increases.

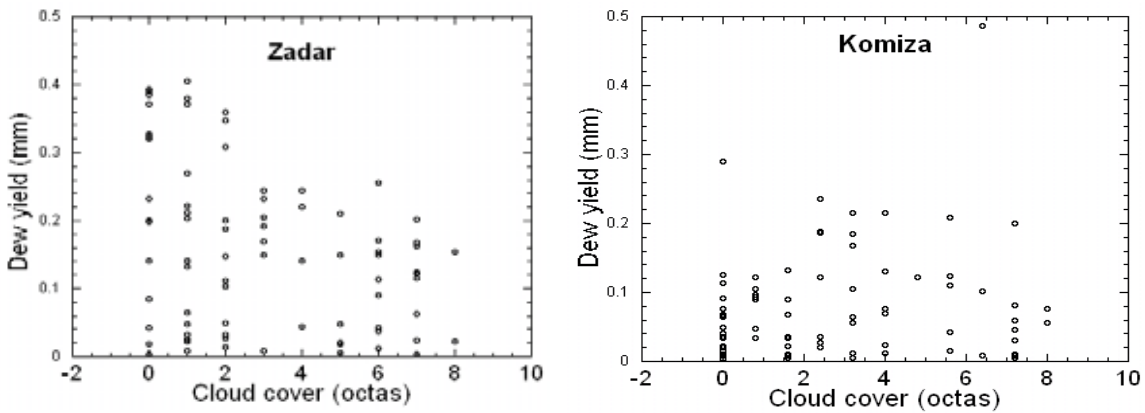

Figure 6. Dew yield with respect to cloud cover (in octas).

## Relative humidity

Relative humidity (RH) is a key parameter for dew formation. Figure 7 contains the dew yield with respect to RH at morning collection time. The largest dew yields correspond to the highest humidity and the lowest cooling temperature. In Table 1, the average humidities for dew formation are close to 78 % for Zadar and Komiža. It is noticeably smaller for Biševo (74%), the reason is presumably the conjunction of more efficient cooling and collecting device and the season under study (summer). In Fig. 7 the dew threshold in RH is about 60% for both Zadar and Komiža; it corresponds to a cooling effect $T-T_a \approx -8°C$. For Biševo, the threshold is noticeably smaller (45 %), corresponding (see Beysens et al., 2005) to a larger cooling $T-T_a \approx -12 °C$. This larger cooling is in agreement with a better condensing efficiency in Biševo, as outlined above.

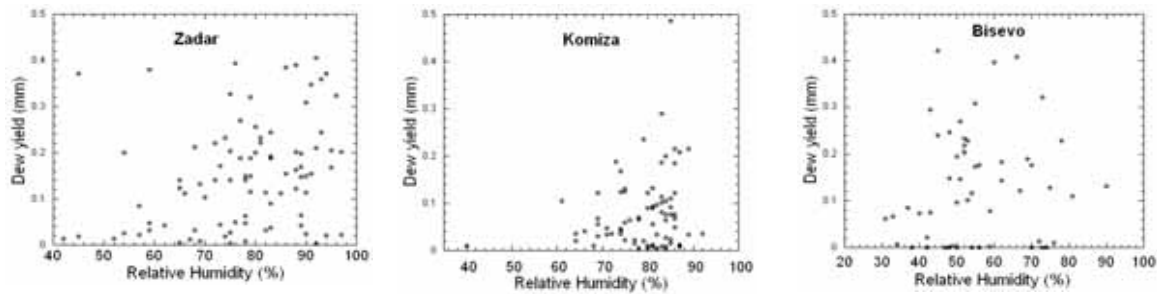

Figure 7. Dew yield with respect to relative humidity.

## Conclusion

The fog water collection on the mountain Velebit shows that there is a great potential of atmospheric water resource.
Maximum amounts of fog water occur in autumn (October and November) together with maximum precipitation amounts caused by cyclone activity over Mediterranean and Adriatic sea.

These promising results on the mountain near the Adriatic cost represent opportunities to obtain a new water resource as an input for the restoration of the degraded vegetation after a forest fire. Some of these areas have suffered from multiple forest fires. Fog water can be intercepted for domestic purposes as on the meteorological station Zavizan located in the Northern Velebit National park where many visitors come every year.

The study on dew collection was motivated by the need to evaluate whether dew water could give a significant amount of water during the dry, summer season in islands and coastal area by the Adriatic sea. From the above investigation on three different sites, it appears that dew water is worth of harvesting, provided that the site is satisfactory (Komiža, due to its particular location, is not) and the condensing material is adequate (thermally isolated, hydrophilic, on a slope). Within these conditions, it appears that dew water can indeed supply to the population an appreciable complement, particularly interesting during the hot, dry season.

The general features of dew formation in the described Dalmatian sites show many similarities. Dew can form under relatively high wind speeds and, with the noticeable exception of the Komiža site, too exposed to ground radiation heating, high dew yields are found in summer. This is especially important for dew collection applications since it is the hot, dry summers that experience moisture stress. This study will continue over several years.

## Acknowledgments

We are especially indebted to A. Vidovic and A. Vidovic for the Zadar data, I. Vitalvić and A. Martinis for the Komiža data and Z. Joncic for the Biševo data.